\documentclass[12pt]{iopart} 
\usepackage{graphicx}
\usepackage{subfig}

\begin{document}

\title{Kleinberg navigation on anisotropic lattices}   

\author{J M Campuzano, J P Bagrow and D ben-Avraham} 
\address{Department of Physics, Clarkson University, Potsdam NY 13699-5820}
\ead{benavraham@clarkson.edu}

\begin{abstract} 
 We study the Kleinberg problem of navigation in Small World networks when the underlying lattice is stretched along a preferred direction.  Extensive simulations confirm that maximally efficient navigation is attained when the length $r$ of long-range links is taken from the distribution $P({\bf r})\sim r^{-\alpha}$, when the exponent $\alpha$ is equal to $2$, the dimension of the underlying lattice, regardless of the amount of anisotropy, but only in the limit of infinite lattice size,
 $L\to\infty$.  For finite size lattices we find an optimal $\alpha(L)$ that depends strongly on $L$.  
 The convergence to $\alpha=2$  as $L\to\infty$ shows interesting power-law dependence on the anisotropy strength.
 \end{abstract}


\date{\today}

\section*{Introduction}

The small-world phenomenon is one of the most intriguing properties of human society.  This describes the fact that unrelated people in a society, who are a very large geographic distance apart from one another, tend to be connected by surprisingly short chains of acquaintances.  This phenomenon was hypothesized in 1929 by Hungarian author Frigyes Karinthy~\cite{citeulike:105595,karinthyChains} and was first observed experimentally in the 1960's with sociologist Stanley Milgram's seminal experiments~\cite{milgramSmallWorld} wherein randomly chosen people were selected to mail a letter to an unknown target person, but were only allowed to send the letter to a friend, who would pass the letter along to another friend, etc., until the target was reached.  Successful transmissions took surprisingly few intermediate people,
lending credibility to the turn of phrase `six degrees of separation,' popularized by Karinthy.   Understanding this phenomenon is an important sociological problem.

To study the underlying mechanism that led to Milgram's results, computer scientist Jon Kleinberg modeled a society as follows~\cite{Kleinberg:SmallWorld_Nature:2000, kleinberg:SmallWorldAlgorithmic:2000}.  Begin with a large, regular square $L\times L$ lattice.  Each node is connected to its nearest lattice neighbors and to a single random node a large distance away.  The probability of nodes $i$ and $j$ being connected by such a long range contact is 
\begin{equation}
	P_{ij}(\alpha) = r_{ij}^{-\alpha} / \sum_{k\neq i} r_{ik}^{-\alpha},
	\label{eqn:kleinbergNavigationProb}
\end{equation}
 where $r_{ij}$ is the Euclidian distance between the two nodes and the sum runs over all nodes in the network except $i$.  Physically, the local lattice connections represent associations with immediate neighbors, fellow townspeople, etc., while long-range contacts might model friends or relatives in another city or country.

We seek to pass the message from a random starting node $s$ to a random target $t$. Of great importance is the fact that each node has no information beyond the locations of its contacts and the target node $t$, so the operational algorithm must be {\it local\/} in character.  Kleinberg has proved that no local algorithm can do better, functionally, than the {\it greedy\/} algorithm~\cite{Kleinberg:SmallWorld_Nature:2000}:  Each message holder passes the message along to whichever of its contacts is closest to $t$, until the message reaches the target.  Moreover, for $\alpha\neq2$ the delivery time $T$ (number of intermediate steps) scales as a {\it power\/} of $L$, while small world behavior and the weakest dependence on lattice size emerges for $\alpha=2$, where $T\sim\ln^2L$ .

The distribution of nodes on a regular square  lattice is too rigid, failing to mimic important features of actual distributions of populations (or computer routers, etc.).  In an effort to account for these we have studied Kleinberg navigation in fractals~\cite{robersonBenAvraham:KleinbergFractals:2006}, showing that the optimal long-contact exponent is then $\alpha=d_{\rm f}$, the fractal dimension of the lattice.  In this letter we study the effects of anisotropy --- another commonly encountered distortion of the ideal Kleinberg lattice.  Our results indicate
that in the limit of lattice size $L\to\infty$ the optimal contact exponent for two-dimensional lattices is still
$\alpha=2$.  For finite $L$, we find an optimal exponent $\alpha(L)$ quite different from the infinite limit.
 The convergence to $\alpha=2$  as $L\to\infty$ shows interesting power-law dependence on the anisotropy strength.

\section*{Anisotropic lattices}

We wish to study the isolated effect of anisotropy on Kleinberg navigation.  To do this, we begin with a regular square lattice ($d=2$) and introduce one of two forms of anisotropy.
\begin{itemize} 
  \item \textbf{Lattice Anisotropy:}  The underlying lattice is stretched horizontally, along the $x$-axis, by a factor $b>0$, such that the area of each cell goes from $1 \times 1$ to $b \times 1$. 
  
  \item \textbf{Angular Anisotropy:}  Long-range contacts are chosen preferably along the vertical direction, by
  a factor $b>0$.  More precisely, the random angle $\theta$ of each long-range contact vector (measured counter-clockwise, from the $x$-axis) is modified to $\theta'$:
  \begin{equation}
  	\theta^\prime = \arctan\left( b \tan\theta  \right).
	\label{eqn:anisoAngleDef}
  \end{equation}  
 
\end{itemize}
Both of these types of anisotropy tend to favor connections in the vertical $y$-direction, if $b>1$, and along
the $x$-direction, if $0<b<1$.

\section*{Simulations}

To simulate Kleinberg navigation efficiently, we use several tricks and approximations.  First, rather than testing
a finite-size square $L\times L$ lattice, we consider an infinite lattice and place the source and target at distance $L$
from one another.  Since  the message always progresses toward the target, by the greedy algorithm,  the message holder remains within a disc of radius $L$ centered on the target node, so in practice only a finite  
number of sites would be explored anyways.  Second, the computation of the normalizing sum in the denominator of Eq.~(\ref{eqn:kleinbergNavigationProb}) is dependent (in finite lattices) upon the location of the node $i$, and can be 
very time consuming.  The infinite lattice circumvents this problem, as the normalizing constant is the same for all
nodes.  Note, however, that $\sum_k r_{ik}^{-\alpha}$ does not converge for $\alpha<d$.
In that case, we imagine a lattice larger than the $L$-disc of activity, with periodic boundary conditions, such that
the normalizing factor is still the same for all sites.  

Because of the monotonic progression toward the target no site is ever revisited in the process.  Moreover, as observed by Kleinberg~\cite{kleinberg:SmallWorldAlgorithmic:2000}, one can think of the long-contact link out of node $i$ as being created at the very instant that the message arrives at $i$.  Thus, the full lattice is unnecessary, and we need keep in memory only the current location of the message holder (and the location of the target). When the message arrives at $i$, we create a random long contact, compare the distances of all five neighbors of $i$ (the four lattice neighbors and the long contact) to the target, and move the message to the site closest to the target.

The long contact is created by choosing a random distance and angle, $(r,\theta)$.  In order to reproduce the correct $P(r_{ij})\sim r_{ij}^{-\alpha}$ the distance $r$ is taken from the distribution $P(r)\sim r^{-\alpha-1}$, to account for the
linear growth of the area of the ring where the contact might fall.  The angle $\theta$ is distributed uniformly between
0 and $2\pi$.  In the case of angular anisotropy, $\theta$ is replaced by $\theta'$, according to~(\ref{eqn:anisoAngleDef}).  Finally, a vector $(r,\theta)$ is drawn from site $i$ and the contact is placed on the site $j$ closest to the vector tip.  

Because of the anisotropy, the angular displacement from the source to the target makes
a difference.  It is sufficient to test only the two extremes of $\theta=0$ and $\theta=\pi/2$ where the target is either parallel or perpendicular to the anisotropy direction.  We note, however, that anisotropy strength $b$ and a target
at $\theta=0$ is equivalent to anisotropy $1/b$ and target at $\theta=\pi/2$.  For this reason, we simply set
the source and target at $(0,0)$ and $(L,0)$, respectively, throughout, and let $b$ vary both below and above the isotropic
divide of $b=1$.

\begin{figure} 
	\subfloat[][Lattice]{
\begingroup%
\makeatletter%
\newcommand{\GNUPLOTspecial}{%
  \@sanitize\catcode`\%=14\relax\special}%
\setlength{\unitlength}{0.0500bp}%
\begin{picture}(4200,3528)(100,0)%
	\footnotesize
  \includegraphics{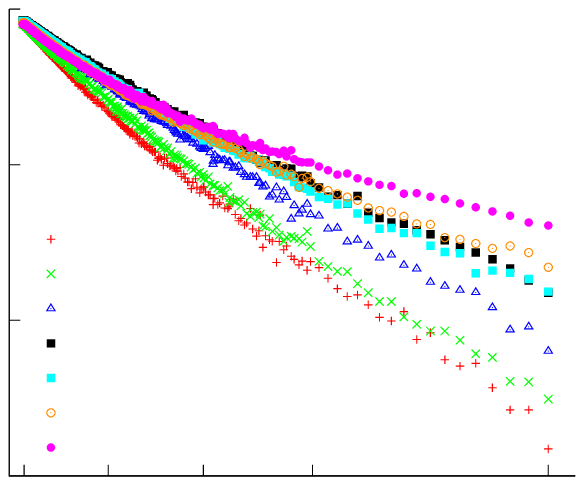}
  \put(1260,763){\makebox(0,0)[l]{\strut{}$b = 32$}}%
  \put(1260,963){\makebox(0,0)[l]{\strut{}$b = 8$}}%
  \put(1260,1163){\makebox(0,0)[l]{\strut{}$b = 2$}}%
  \put(1260,1363){\makebox(0,0)[l]{\strut{}$b = 1$}}%
  \put(1260,1563){\makebox(0,0)[l]{\strut{}$b = 1/2$}}%
  \put(1260,1763){\makebox(0,0)[l]{\strut{}$b = 1/8$}}%
  \put(1260,1963){\makebox(0,0)[l]{\strut{}$b = 1/32$}}%
  \put(2530,100){\makebox(0,0){\strut{}$L$}}%
  \put(200,1944){%
  \makebox(0,0){\strut{}$\alpha_{\mathrm{min}}$}%
  }%
  \put(4005,400){\makebox(0,0){\strut{}$10^3$}}%
  \put(2647,400){\makebox(0,0){\strut{}$10^4$}}%
  \put(2018,400){\makebox(0,0){\strut{}$10^5$}}%
  \put(1470,400){\makebox(0,0){\strut{}$10^7$}}%
  \put(986,400){\makebox(0,0){\strut{}$10^{18}$}}%
  \put(780,3288){\makebox(0,0)[r]{\strut{} 2}}%
  \put(780,2392){\makebox(0,0)[r]{\strut{} 1.9}}%
  \put(780,1496){\makebox(0,0)[r]{\strut{} 1.8}}%
  \put(780,600){\makebox(0,0)[r]{\strut{} 1.7}}%
\end{picture}%
\endgroup

		\label{fig:latticeAnisoData}
	}
	\subfloat[][Angle]{
\begingroup%
\makeatletter%
\newcommand{\GNUPLOTspecial}{%
  \@sanitize\catcode`\%=14\relax\special}%
\setlength{\unitlength}{0.0500bp}%
\begin{picture}(4200,3528)(100,0)%
	\footnotesize
  \includegraphics{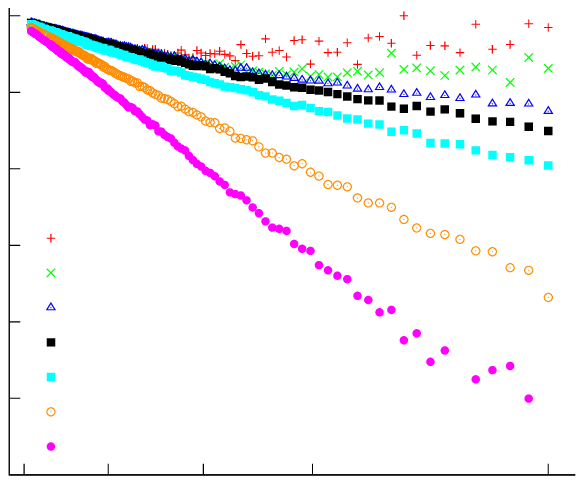}
  \put(1260,763){\makebox(0,0)[l]{\strut{}$b = 32$}}%
  \put(1260,963){\makebox(0,0)[l]{\strut{}$b = 8$}}%
  \put(1260,1163){\makebox(0,0)[l]{\strut{}$b = 2$}}%
  \put(1260,1363){\makebox(0,0)[l]{\strut{}$b = 1$}}%
  \put(1260,1563){\makebox(0,0)[l]{\strut{}$b = 1/2$}}%
  \put(1260,1763){\makebox(0,0)[l]{\strut{}$b = 1/8$}}%
  \put(1260,1963){\makebox(0,0)[l]{\strut{}$b = 1/32$}}%
  \put(2530,100){\makebox(0,0){\strut{}$L$}}%
  \put(200,1944){%
  \makebox(0,0){\strut{}$\alpha_{\mathrm{min}}$}%
  }%
  \put(4005,400){\makebox(0,0){\strut{}$10^3$}}%
  \put(2647,400){\makebox(0,0){\strut{}$10^4$}}%
  \put(2018,400){\makebox(0,0){\strut{}$10^5$}}%
  \put(1470,400){\makebox(0,0){\strut{}$10^7$}}%
  \put(986,400){\makebox(0,0){\strut{}$10^{18}$}}%
  \put(780,3244){\makebox(0,0)[r]{\strut{} 2}}%
  \put(780,2803){\makebox(0,0)[r]{\strut{} 1.9}}%
  \put(780,2363){\makebox(0,0)[r]{\strut{} 1.8}}%
  \put(780,1922){\makebox(0,0)[r]{\strut{} 1.7}}%
  \put(780,1481){\makebox(0,0)[r]{\strut{} 1.6}}%
  \put(780,1041){\makebox(0,0)[r]{\strut{} 1.5}}%
  \put(780,600){\makebox(0,0)[r]{\strut{} 1.4}}%
\end{picture}%
\endgroup
		\label{fig:angleAnisoData}
	}
   \caption{ Simulations for lattice and angle anisotropy.  A horizontal scale of $1/\ln^2L$ is used throughout.  All curves approach $\alpha(\infty)$, regardless of $b$.  For the lattice case, there is a crossover effect where curves for $b>1$ dip below the $b=1$ curve.  For the angular case, curves for $b>1$ approach the infinite limit at differing rates, while curves for $b<1$ evetually collapse onto the $b=1$ curve.  These phenomena are further explored in Fig.~\protect{\ref{fig:renormedCubics}}. See Fig.~\protect{\ref{fig:accuracyExtrap}} for the extrapolated $\alpha(\infty)$.   }
   \label{fig:AnisoData}
\end{figure}

\begin{figure}[h]
	\subfloat[][Lattice]{
\begingroup%
\makeatletter%
\newcommand{\GNUPLOTspecial}{%
  \@sanitize\catcode`\%=14\relax\special}%
\setlength{\unitlength}{0.0500bp}%
\begin{picture}(4356,3049)(0,0)%
	\footnotesize
  \includegraphics{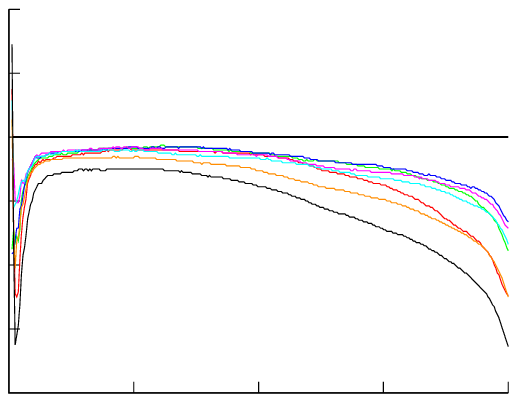}
  \put(2578,100){\makebox(0,0){\strut{}Fraction of data fitted, from left}}%
  \put(200,1704){%
  \makebox(0,0){\strut{}$\alpha_{\mathrm{min}}(\infty)$}%
  }%
  \put(4016,400){\makebox(0,0){\strut{} 1}}%
  \put(3297,400){\makebox(0,0){\strut{} 0.75}}%
  \put(2578,400){\makebox(0,0){\strut{} 0.5}}%
  \put(1859,400){\makebox(0,0){\strut{} 0.25}}%
  \put(1140,400){\makebox(0,0){\strut{} 0}}%
  \put(1020,2809){\makebox(0,0)[r]{\strut{} 2.008}}%
  \put(1020,2441){\makebox(0,0)[r]{\strut{} 2.004}}%
  \put(1020,2073){\makebox(0,0)[r]{\strut{} 2}}%
  \put(1020,1705){\makebox(0,0)[r]{\strut{} 1.996}}%
  \put(1020,1336){\makebox(0,0)[r]{\strut{} 1.992}}%
  \put(1020,968){\makebox(0,0)[r]{\strut{} 1.988}}%
  \put(1020,600){\makebox(0,0)[r]{\strut{} 1.984}}%
\end{picture}%
\endgroup

	}
	\subfloat[][Angle]{
\begingroup%
\makeatletter%
\newcommand{\GNUPLOTspecial}{%
  \@sanitize\catcode`\%=14\relax\special}%
\setlength{\unitlength}{0.0500bp}%
\begin{picture}(4356,3049)(0,0)%
	\footnotesize
  \includegraphics{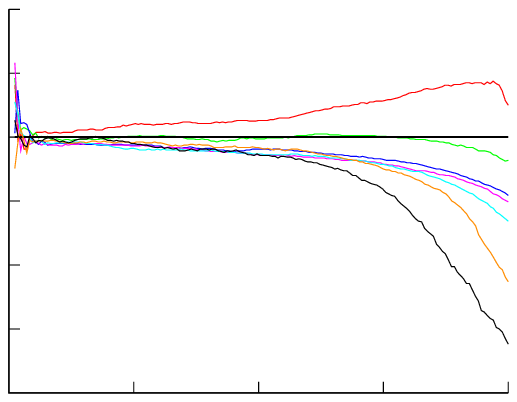}
  \put(2578,100){\makebox(0,0){\strut{}Fraction of data fitted, from left}}%
  \put(200,1704){%
  \makebox(0,0){\strut{}$\alpha_{\mathrm{min}}(\infty)$}%
  }%
  \put(4016,400){\makebox(0,0){\strut{} 1}}%
  \put(3297,400){\makebox(0,0){\strut{} 0.75}}%
  \put(2578,400){\makebox(0,0){\strut{} 0.5}}%
  \put(1859,400){\makebox(0,0){\strut{} 0.25}}%
  \put(1140,400){\makebox(0,0){\strut{} 0}}%
  \put(1020,2809){\makebox(0,0)[r]{\strut{} 2.016}}%
  \put(1020,2441){\makebox(0,0)[r]{\strut{} 2.008}}%
  \put(1020,2073){\makebox(0,0)[r]{\strut{} 2}}%
  \put(1020,1705){\makebox(0,0)[r]{\strut{} 1.992}}%
  \put(1020,1336){\makebox(0,0)[r]{\strut{} 1.984}}%
  \put(1020,968){\makebox(0,0)[r]{\strut{} 1.976}}%
  \put(1020,600){\makebox(0,0)[r]{\strut{} 1.968}}%
\end{picture}%
\endgroup

	}
	\caption{Extrapolation results for lattice~(a) and angular~(b) anisotropy.  Extrapolating to $1/\ln^2 L \to 0$ with a linear least squares fit to the curves in Fig.~\protect{\ref{fig:AnisoData}} shows excellent convergence of $\alpha(\infty )$ to the expected value of $d =2$.  Good values should occur when the curves are flattest, which happens roughly around 0.25.  A more robust fitting procedure could be used, but the accuracy of these results imply that it is unnecessary.   The horizontal lines at $\alpha = 2$ provides a guide for the eye.\label{fig:accuracyExtrap} }%
\end{figure}

Simulations were performed for various values of $b$ over a large range of $\alpha$ and $L$, each averaged 1000 times.  For each $b$ and $L$, the minimum $\alpha$ was computed by first fitting a fifth-order polynomial\footnote{A parabola could be fitted to the data closest to the minimum, but we must first know what is `closest.'   A higher-order polynomial overcomes this difficulty, similar to including higher order terms in a series expansion near the minmum of a function.} to the averaged data, then using Newton's Method on the polynomial's derivative.  Finally $\alpha_{\min}$ was plotted as a function of $1/\ln^2 L$ for each chosen value of $b$.  These are shown in Fig.~\ref{fig:AnisoData} and indicate that $\alpha_{\min} \to 2$ as $L \to \infty$, regardless of $b$. In Fig.~\ref{fig:accuracyExtrap} we show detailed results of the extrapolation to $L\to\infty$.

\begin{figure}
   \centering
	\subfloat[][Lattice]{
\begingroup%
\makeatletter%
\newcommand{\GNUPLOTspecial}{%
  \@sanitize\catcode`\%=14\relax\special}%
\setlength{\unitlength}{0.0500bp}%
\begin{picture}(4248,3276)(0,0)%
	\footnotesize
  \includegraphics{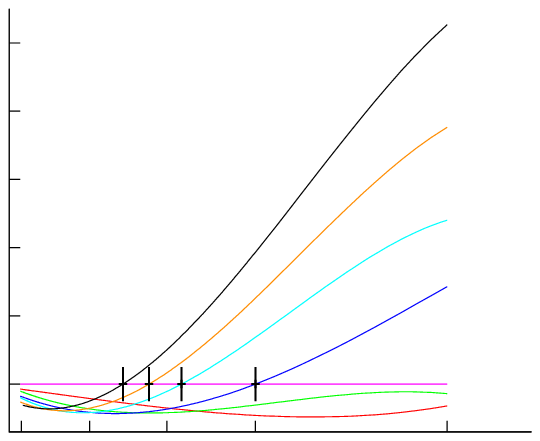}
  \put(3487,2945){\makebox(0,0)[l]{\strut{}\small $b = 64$}}%
  \put(3487,2354){\makebox(0,0)[l]{\strut{}\small $b = 32$}}%
  \put(3487,1819){\makebox(0,0)[l]{\strut{}\small $b = 16$}}%
  \put(3487,1436){\makebox(0,0)[l]{\strut{}\small $b = 8$}}%
  \put(3487,849){\makebox(0,0)[l]{\strut{}\small $b = 4$}}%
  \put(3487,680){\makebox(0,0)[l]{\strut{}\small $b = 2$}}%
  \put(2404,100){\makebox(0,0){\strut{}$L$}}%
  \put(200,1818){%
  \makebox(0,0){\strut{}cubic fit $p_b - p_1$}%
  }%
  \put(3422,400){\makebox(0,0){\strut{}$10^3$}}%
  \put(2318,400){\makebox(0,0){\strut{}$10^4$}}%
  \put(1808,400){\makebox(0,0){\strut{}$10^5$}}%
  \put(1363,400){\makebox(0,0){\strut{}$10^7$}}%
  \put(970,400){\makebox(0,0){\strut{}$10^{18}$}}%
  \put(780,2840){\makebox(0,0)[r]{\strut{}0.05}}%
  \put(780,2447){\makebox(0,0)[r]{\strut{}0.04}}%
  \put(780,2054){\makebox(0,0)[r]{\strut{}0.03}}%
  \put(780,1661){\makebox(0,0)[r]{\strut{}0.02}}%
  \put(780,1268){\makebox(0,0)[r]{\strut{}0.01}}%
  \put(780,875){\makebox(0,0)[r]{\strut{}0}}%
\end{picture}%
\endgroup

	}
	\subfloat[][Angle]{
\begingroup%
\makeatletter%
\newcommand{\GNUPLOTspecial}{%
  \@sanitize\catcode`\%=14\relax\special}%
\setlength{\unitlength}{0.0500bp}%
\begin{picture}(4248,3276)(0,0)%
	\footnotesize
  \includegraphics{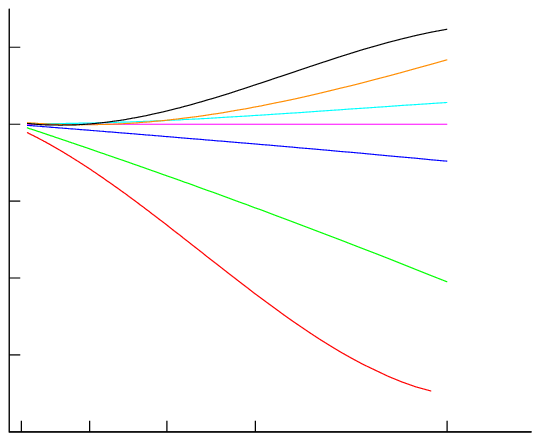}
  \put(3403,835){\makebox(0,0)[l]{\strut{}\small $b = 32$}}%
  \put(3487,1464){\makebox(0,0)[l]{\strut{}\small $b = 8$}}%
  \put(3487,2159){\makebox(0,0)[l]{\strut{}\small $b = 2$}}%
  \put(3487,2919){\makebox(0,0)[l]{\strut{}\small $b = 1/32$}}%
  \put(3487,2743){\makebox(0,0)[l]{\strut{}\small $b = 1/8$}}%
  \put(3487,2497){\makebox(0,0)[l]{\strut{}\small $b = 1/2$}}%
  \put(2404,100){\makebox(0,0){\strut{}$L$}}%
  \put(200,1818){%
  \makebox(0,0){\strut{}cubic fit $p_b - p_1$}%
  }%
  \put(3422,400){\makebox(0,0){\strut{}$10^3$}}%
  \put(2318,400){\makebox(0,0){\strut{}$10^4$}}%
  \put(1808,400){\makebox(0,0){\strut{}$10^5$}}%
  \put(1363,400){\makebox(0,0){\strut{}$10^7$}}%
  \put(970,400){\makebox(0,0){\strut{}$10^{18}$}}%
  \put(780,2815){\makebox(0,0)[r]{\strut{} 0.1}}%
  \put(780,2372){\makebox(0,0)[r]{\strut{} 0}}%
  \put(780,1929){\makebox(0,0)[r]{\strut{}-0.1}}%
  \put(780,1486){\makebox(0,0)[r]{\strut{}-0.2}}%
  \put(780,1043){\makebox(0,0)[r]{\strut{}-0.3}}%
  \put(780,600){\makebox(0,0)[r]{\strut{}-0.4}}%
\end{picture}%
\endgroup

	}
   \caption{ Curves relative to the isotropic divide $b=1$, for the lattice~(a) and angular~(b) case.  To provide a measure of smoothing, cubic polynomials $p_b$ were fitted to the curves in Fig.~\protect{\ref{fig:AnisoData}}.  To clarify the impact of anisotropy, we show the behavior relative to the isotropic case, by subtracting $p_1$ from each $p_b$.  This maps the isotropic curve to a horizontal line and introduces only minor distortion.  The crossover behavior for $b>1$ is clearly displayed.  
   }
   \label{fig:renormedCubics}
\end{figure}

To further clarify the behavior shown in Fig.~\ref{fig:AnisoData}, the following procedure was performed.  First fit a cubic polynomial $p_b$, using least squares, to each $b$'s curve.  Then, subtract that polynomial from the isotropic case, $p_b - p_1$.  This maps $b=1$ to the horizontal axis and gives the behavior of the $b \neq 1$ curves ``relative'' to the isotropic curve.  These are shown in Fig.~\ref{fig:renormedCubics}.  The different behavior for each type of anisotropy is clear:  for both types of anisotropy the results for $b>1$ show dramatic differences from the  isotropic case of $b=1$ (the differences for $b<1$ and large $L$ are negligible). For lattice anisotropy, the $b>1$ curves start above the $b=1$ curve and cross below until they eventually converge at a similar rate, as $L\to\infty$.  For the angular anisotropy, the $b>1$ curves approach $\alpha(\infty)$ at a different rate than the $b=1$ curve, resulting in distinctly different slopes in the plots of Fig.~\ref{fig:angleAnisoData}.

\begin{figure} 
   \centering
   	\subfloat[][Lattice]{
\begingroup%
\makeatletter%
\newcommand{\GNUPLOTspecial}{%
  \@sanitize\catcode`\%=14\relax\special}%
\setlength{\unitlength}{0.0500bp}%
\begin{picture}(4320,3024)(0,0)%
	\footnotesize
  \includegraphics{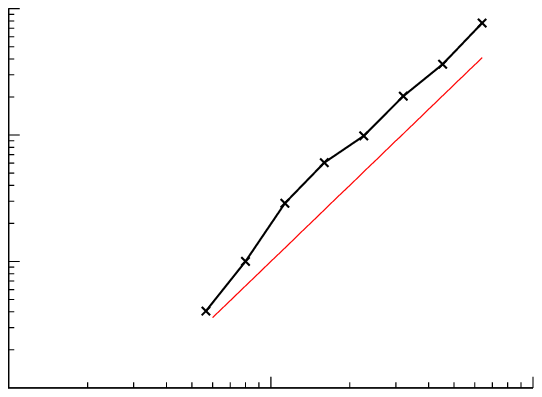}
  \put(2470,100){\makebox(0,0){\strut{}$b$}}%
  \put(260,1692){%
  \makebox(0,0){\strut{}$L_{\mathrm{crossover}}$}%
  }%
  \put(3980,400){\makebox(0,0){\strut{} 100}}%
  \put(2470,400){\makebox(0,0){\strut{} 10}}%
  \put(960,400){\makebox(0,0){\strut{} 1}}%
  \put(840,2784){\makebox(0,0)[r]{\strut{}$10^{6}$}}%
  \put(840,2056){\makebox(0,0)[r]{\strut{}$10^{5}$}}%
  \put(840,1328){\makebox(0,0)[r]{\strut{}$10^{4}$}}%
  \put(840,600){\makebox(0,0)[r]{\strut{}$10^{3}$}}%
\end{picture}%
\endgroup

	}
	\subfloat[][Angle]{
\begingroup%
\makeatletter%
\newcommand{\GNUPLOTspecial}{%
  \@sanitize\catcode`\%=14\relax\special}%
\setlength{\unitlength}{0.0500bp}%
\begin{picture}(4320,3024)(0,0)%
	\footnotesize
  \includegraphics{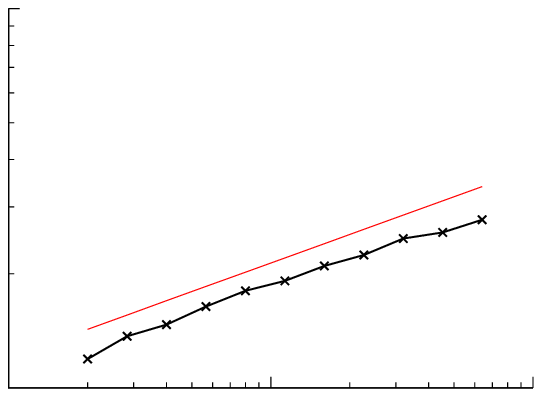}
  \put(2470,100){\makebox(0,0){\strut{}$b$}}%
  \put(260,1692){%
  \makebox(0,0){\strut{}$\left| \mathrm{slope} \right|$}%
  }%
  \put(3980,400){\makebox(0,0){\strut{} 100}}%
  \put(2470,400){\makebox(0,0){\strut{} 10}}%
  \put(960,400){\makebox(0,0){\strut{} 1}}%
  \put(840,2784){\makebox(0,0)[r]{\strut{}$10^{2}$}}%
  \put(840,600){\makebox(0,0)[r]{\strut{}$10^{1}$}}%
\end{picture}%
\endgroup

	}
\caption{ Dependence of results on anisotropy strength $b$ for lattice~(a) and angular~(b) anisotropy.  The straight lines are of slope 2 and $1/4$, respectively.  \label{fig:loglogLatticeCrossover}}
   
\end{figure}

The observed ``crossover'' behavior present in the lattice anisotropy is somewhat unexpected.  The crossover point, $L_\mathrm{crossover}(b)$,  is explored by finding the zero of each $p_b - p_1$.  These are plotted in Fig.~\ref{fig:loglogLatticeCrossover}a, and seem to indicate a power law relationship, $L_\mathrm{crossover}(b)\sim b^2$.  Likewise, the different slopes for angular anisotropy, plotted in Fig.~\ref{fig:loglogLatticeCrossover}b, show power-law behavior and seem to increase roughly as $b^{1/4}$.  What is responsible for these phenomena remains an open question. 

\section*{Conclusions}

We have shown, by extensive numerical simulations, that Kleinberg navigation in two-dimensional lattices
with two types of anisotropy displays the same gross characteristics as navigation in isotropic lattices.  In particular,
the optimal long-contact exponent in the limit of infinitely distant source and target remains $\alpha=2$, even in the
presence of anisotropy.

It is worthwhile to note that the actual values for the optimal exponent $\alpha(L)$
for finite $L$ can differ considerably from the limit $\alpha=2$, even for reasonably large lattices.  Thus, for 
practical applications the optimal exponent ought to be evaluated on a case by case basis.

The modes of convergence to the limit $L\to\infty$ show intriguing power-law dependence upon the strength
of the anisotropy.  A theoretical explanation for this behavior remains the subject of future work.

\ack
We gratefully acknowledge NSF award PHY0555312 for partial funding (DbA), and the support of a NSF Graduate Research Fellowship (JPB).  JMC has been supported as a summer research undergraduate student by the 
McNair program at Carkson University.

\end{document}